\documentclass{article}

\usepackage{arxiv}

\usepackage[utf8]{inputenc} 
\usepackage[T1]{fontenc}    
\usepackage{hyperref}       
\usepackage{url}            
\usepackage{booktabs}       
\usepackage{amsfonts}       
\usepackage{nicefrac}       
\usepackage{microtype}      
\usepackage{lipsum}
\usepackage{graphicx}

\title{Quantifying Latent Moral Foundations in Twitter Narratives: The Case of the Syrian White Helmets Misinformation}

\author{
  Ece Çiğdem Mutlu, Toktam Oghaz, Ege Tütüncüler, Jasser Jasser, Ivan Garibay \\
    Complex Adaptive System Laboratory\\
  University of Central Florida\\
  Orlando, FL 32816 \\
  \texttt{ece.mutlu, toktam.oghaz, ege.tutunculer, jasser.jasser, igaribay@ucf.edu} \\
}

\begin{document}
\maketitle

\begin{abstract}
For years, many studies employed sentiment analysis to understand the reasoning behind people's choices and feelings, their communication styles, and the communities which they belong to. We argue that gaining more in-depth insight into moral dimensions coupled with sentiment analysis can potentially provide superior results. Understanding moral foundations can yield powerful results in terms of perceiving the intended meaning of the text data, as the concept of morality provides additional information on the unobservable characteristics of information processing and non-conscious cognitive processes. Therefore, we studied latent moral loadings of Syrian White Helmets-related tweets of Twitter users from April 1st, 2018 to April 30th, 2019. For the operationalization and quantification of moral rhetoric in tweets, we use Extended Moral Foundations Dictionary in which five psychological dimensions (Harm/Care, Fairness/Reciprocity, In-group/Loyalty, Authority/Respect and Purity/Sanctity) are considered. We show that people tend to share more tweets involving the virtue moral rhetoric than the tweets involving the vice rhetoric. We observe that the pattern of the moral rhetoric of tweets among these five dimensions are very similar during different time periods, while the strength of the five dimension is time-variant. Even though there is no significant difference between the use of Fairness/Reciprocity, In-group/Loyalty or Purity/Sanctity rhetoric, the less use of Harm/Care rhetoric is significant and remarkable. Besides, the strength of the moral rhetoric and the polarization in morality across people are mostly observed in tweets involving Harm/Care rhetoric despite the number of tweets involving the Harm/Care dimension is low.
\end{abstract}


\section{Introduction}
For thousands of years, humankind has pondered the question of morality. Ancient philosophers included several conventional notions in their definitions of what is moral; which include Socrates’, Plato’s and Aristotle’s emphasis on happiness (eudaimonia), virtues, moderation, and justice \cite{parry2004ancient}. Understanding such notions would, in the eyes of ancient and modern philosophers alike, provide a reflection of the human activity in pursuing their goals and behaving in certain ways. To this end, modern scientists have produced numerous theories to grasp the role of morality in people’s cognitive processes and behavioral functioning in society. It is essential to understand the differences in the notion of morality at both the cultural and individual levels, as morality guides human social interactions and can potentially lead to polarity, violence, and hostility when there is a clash of moral values within a society \cite{graham2013moral,garibay2019polarization}.

Differences in morality can result in polarity within a social group and can fuel societal tensions caused by disinformation efforts of malicious agents. The unintermediated flow of information in online social media has paved the way for the widespread use of adversarial narratives, which can be defined as intentionally distributed disinformation narratives with the aim of dividing internet users and inflaming social tensions by exploiting their moral values and foundations \cite{AdvNar}. Thus, identifying the bits-and-pieces of morality components in social media content, such as user tweets revolving around certain adversarial narratives in Twitter, can help combat the spread of misinformation and the social engineering efforts of adversaries. Furthermore, gaining more in-depth insight into morality dimensions can potentially provide superior results compared to what the extant literature achieves by using sentiment analyses in combating disinformation \cite{rezapour2019enhancing}. Understanding moral foundations can yield powerful results in terms of perceiving the intended meaning of the text data, as the concept of morality provides additional information on the unobservable characteristics of information processing and non-conscious cognitive processes. Considering that moral values vary significantly across cultures and yet many recurrent themes are observed and that each culture builds its societal and ideological narratives on top of its moral virtues, an enhanced understanding of morality can prove to be a valuable tool in deterring disinformation narratives by adversaries. 

Sharing information, interests, and opinions of people in social media platforms has led scientists from various fields to focus on user-generated text data. The text data is extensively used to study, analyze, and extract people's cultural values, behavior, opinions, and emotions \cite{mutlu2020degree}. Easy access to these platforms, being able to reach multiple people simultaneously, and the free self-expression dynamics within social media have increased the popularity of research that relies on these platforms. Many psychological and cognitive science-related studies try to understand the underlying beliefs, attitudes, and emotions which drive human actions \cite{reisenzein2009emotions,whitley2008cognition,paulus2012emotion}.

While making analyses on text data has been a common application in the literature, the availability of large-scale time-series text data from social media platforms has made such studies more rigorous and diverse. However, supervised learning techniques struggle to address these applications due to the challenge of finding annotated data for human utterances. Oftentimes, even humans themselves are not aware of the latent factors that guide their selection of specific language use to express opinions and attitudes. Therefore, many theory-driven studies have benefited from the psychological dictionaries coupled with data-driven natural language processing methods. A commonly used approach is the word-count method, which is pioneered by the LIWC package \cite{pennebaker2001linguistic} and its widespread application of specific dictionaries. 

Using lexicon data to operationalize and identify the fluid and highly subjective concepts of morality may appear questionable at first; however, this approach is not ungrounded and is found to be a reliable method. Weber et al. reports that the overlap between human coders and the lexicon in classifying and identifying morality dimensions ranges from 0.73 to 1.00 in multiple studies \cite{graham2009liberals, hofmann2014morality,feinberg2013moral,clifford2013words}, thereby establishing the lexicon as a fairly reliable identifier of morality dimensions of the Moral Foundations Theory. Indeed, in the past decade, many studies have applied the lexicon approach to analyze moral foundations using text data \cite{kaur2016quantifying,hoover2019moral,sagi2014moral}.

\section{Method}
\subsection{Twitter Data Collection}
The data investigated in this work is provided by Leidos Inc\footnote{\url{https://www.leidos.com}} as part of the "Computational Simulation of Online Social Behavior (SocialSim)" DARPA program\footnote{\url{https://www.darpa.mil/program/computational-simulation-of-online-social-behavior}}. The data consists of 1,052,821 tweets related to the disinformation campaigns carried against the White Helmets from April 1st, 2018 to April 30th, 2019. The narratives included within the content of these tweets are mostly attacks against the integrity of the White Helmets' work and mission statement, accusing the organization of being foreign agents, and nullifying the narrative of the chemical attack by censuring the organization of staging the event \cite{starbird2018ecosystem}.

Each tweet in our data has an identification number (ID) along with its content, and the IDs of the tweets that interacted with. This will help us construct the information cascade associated with our data and trace the argument and discussion that took place in regard to the specific narratives we defined. To build the information cascade (retweet cascade), the tweets $t$ are separated into two main sets, parent nodes set $P$, and child nodes set $C$. The intersection of the two sets contains all the the tweets that are both parents and children at the same time. The roots $t_r$ are tweets that do not have a parent and are the start of the cascades. In this work, we considered the largest 600 cascades, which consists of 365085 tweets.

Every retweet cascade starts with one root $t_r$ which belongs in $P$ but not in $C$ as shown in equation.\ref{eq:root}. Any child of that root that is also a parent of another child (we refer to it as '\textit{parent}' $t_p$) belongs to the intersection of $P$ and $C$, as shown in equation.\ref{eq:parent}. Any tweet that is not a parent is a child $t_c$ only and belongs in $C$ but not in $P$, as shown in equation.\ref{eq:child}. Constructing the chain of interaction based on these node sets would give us a temporal tree structure of the conversation regarding a particular narrative. Every cascade tweet corpus is aggregated for analysis purposes. The aggregated text is then preprocessed using Natural Language Processing (NLP) techniques to clean unwanted keywords, hashtags, URLs, etc. included in the text, as discussed in section \ref{sec:data-preprocessing}. Quantifying the moral foundations based on a lexicon is then implemented on the cleaned corpus, as discussed in Section 2.3.

\begin{equation}
    t_r \in P - C
    \label{eq:root}
\end{equation}
\begin{equation}
    t_p \in P \cap C
    \label{eq:parent}
\end{equation}
\begin{equation}
    t_c \in C - P
    \label{eq:child}
\end{equation}

\subsection{Data Pre-processing}
\label{sec:data-preprocessing}
The Twitter dataset used in this paper comprises of tweet texts and tweet features such as text language; however, topic mining on short-text data sets, including Twitter data set, is challenging as a result of limited word co-occurrence and contextual information. Accordingly, extracting meaningful topics necessitates text aggregation in a way to enhance the text with context and related keywords. To tackle this challenge, we prepared pseudo-documents via aggregating root, parent, and reply/quote/retweet comments for each activity cascade such that the aggregated text data is timely order. Further information on conversation cascade formation is provided in section 2.1. This text aggregation method results in preparing pseudo-documents rich in context and related words. We continued text preprocessing for our topic analysis in cleaning up the data by removing usernames, short URLs, emoticons, as well as punctuation marks. Next, we removed the hashtag symbols from the text data. 

\subsection{Quantifying Moral Foundations in a Text}
In order to operationalize and capture dimensions of morality in our Twitter text data, we draw from social psychology literature and use the Moral Foundations Theory (MFT) \cite{graham2009liberals,haidt2004intuitive}. The moral rhetoric that provides a basis for our analyses can be defined as the linguistic component for expressing various moral concerns by taking a moral stance towards an issue \cite{sagi2014measuring}. MFT contends that five psychological subsystems constitute moral cognition, which manifests themselves as moral concerns or intuitions. Each of these five morality-related psychological components includes dimensions of virtues and vices. Specifically, 
\begin{itemize}
    \item Harm/Care concern is associated with the protection of self and others from the harm's way,
    \item Fairness/Reciprocity concern is related to justice in cooperative acts, prevention of dishonesty, and reciprocity in social interactions,
    \item In-group/Loyalty dimension is based on the expressions of self-sacrifice for both ends of the virtue-vice spectrum, such as patriotism-betrayal, faithfulness-unfaithfulness,
    \item Authority/Respect component expresses concerns related to subordination and respect,
    \item Purity/Sanctity is associated with sanctity in the virtue dimension and degradation and pollution in the vice dimension
\end{itemize}

To quantify the the moral foundations in Twitter text data, we use the extended version of Moral Foundation Dictionary (EMFD) that is discussed in a very recent study \cite{araque2020moralstrength}. Corpus size of the dictionary is given in Table \ref{tab:corpus} for each dimension.  

\begin{table}[h!]
  \centering
  \caption{Corpus Size of The Extended Moral Foundations Dictionary (EMFD)}
  \label{tab:corpus}
  \begin{tabular}{ccl}
    \toprule
   Moral Dimension&Virtue&Vice\\
    \midrule
    Care/Harm & 95 & 85\\
    Fairness/Reciprocity & 69 & 57\\
    Loyalty/Ingroup & 99 & 72\\
    Authority/respect & 160 & 101\\
    Purity/Sanctity & 97 & 161\\
  \bottomrule
\end{tabular}
\end{table}

In order to understand the effects of moral foundations on opinion formation, dissemination, and polarization, this study makes use of user-generated content on Twitter. However, there are multiple challenges that need to be pointed out before we take on this task. First, each tweet can only contain up to 280 characters, which often poses a difficulty for algorithms to fully understand the true meaning of the message, due to users being constrained to writing a short text. Second, the concept of morality is very subjective, and it is a challenging task for the human brain to annotate morality in a data set. It is a formidable task for machine learning techniques, even more so than human annotation methods. Third, multiple morality dimensions can coexist in a tweet since moral foundations do not complement each other, and this brings an extra layer of complexity to the task of clustering the data \cite{garten2016morality}. Finally, there is limited variability in our data set since it focuses on a single topic only. Oftentimes, a multi-topic analysis is sought after in analyzing moral foundations to capture greater diversity across tweets. Our data set falls short in that respect because, in a single topic, we observe a great degree of similarity across tweets, as these tweets often originate from the same root. Some tweets share exactly the same moral foundations because they are retweets; or, similar morality dimensions are observed when they are replies or quotes.

\subsection{Latent Dirichlet Allocation (LDA)}
LDA is one of the most commonly used topic analysis method due to its simplicity and applicability to big data sets. One of the most challenging tuning parameter of this algorithm is defining the number of topic clusters ($K$) with their probable initial proportions. The generative process starts with drawing the topic distribution over the vocabulary $\beta_k \sim Dirichlet(\eta,...,\eta)$ for $k \in \{1,...,K\}$. Suppose that $d \in \{1,...,D\}$ for each document. Then, topic proportions are drawn from $\theta \sim Dirichlet(\alpha,...,\alpha)$ and for each word, $w \in \{1,...,N\}$, topic assignments and words are obtained from $z_{dn} \sim Multinomial(\theta_d)$ and $w_{dn} \sim Multinomial(\beta_{z_{dn}})$, respectively. 

In this study, the stochastic variational inference method is used to optimize LDA. The advantage of stochastic search in the approximate posterior inference comes from the scalability of its method. In each mini batch, the global variational parameter $\lambda_k$ for each topic; and the local variational parameters of topic proportion $\gamma_d$ and multinomial parameter of per-word topic $\phi_d$ are updated as follows: In each iteration $t$, a document $d$ is sampled from the collection and optimal variational parameters of $\gamma_d$ and $\phi_{d,1:N}$ in the local phase. In the global phase, on the other hand, intermediate topics are obtained from:
\begin{equation}
    \hat{\lambda}_k=\eta + D\sum_{n=1}^N \phi^k_{dn} w_{dn}
\end{equation}
Then, in the next iteration intermediate topics are updated as the weighted combination of the intermediate topics in the current iteration and the current topics by: 

\begin{equation}
    {\lambda}_k^{t+1}=(1-\rho_t){\lambda}_k^{t}+\rho_t \hat{\lambda}_k
\end{equation}
where $\rho_t$ denotes the step-size schedule. 

Here, the main purpose of using stochastic search is to optimize the variational objective \cite{hoffman2013stochastic}. Different solver techniques for optimization of LDA algorithm have been tried and best perplexity and coherence values are obtained with the use of stochastic variational Bayes solver \cite{foulds2013stochastic,hoffman2013stochastic} compared to collapsed Gibbs sampling \cite{griffiths2004finding}, approximate variational Bayes \cite{asuncion2012smoothing} and the zeroth order collapsed variational Bayes solvers \cite{teh2007collapsed}. 

\subsection{Cross-Recurrence Quantification Analysis (CRQA)}
CRQA is a nonlinear correlation analysis technique which is used to capture the coupling -togetherness- of two time series. Its superiority to the traditional Pearson correlation analysis comes from its capability of showing temporal co-variation of two time series rather than analyzing them as a whole \cite{coco2014cross}. Having no assumption about linearity and the underlying distribution and being highly robust to the outliers make this method increasingly prominent \cite{wallot2019multidimensional}. Although there are multiple performance measures in CRQA, we used Shannon Entropy as a metric in understanding the similarity in the phase space behavior of different time series, which measures the percentage of the the co-variation of time series. 

\section{Results}
\begin{figure} 
    \centering
    \includegraphics[width=0.6\linewidth]{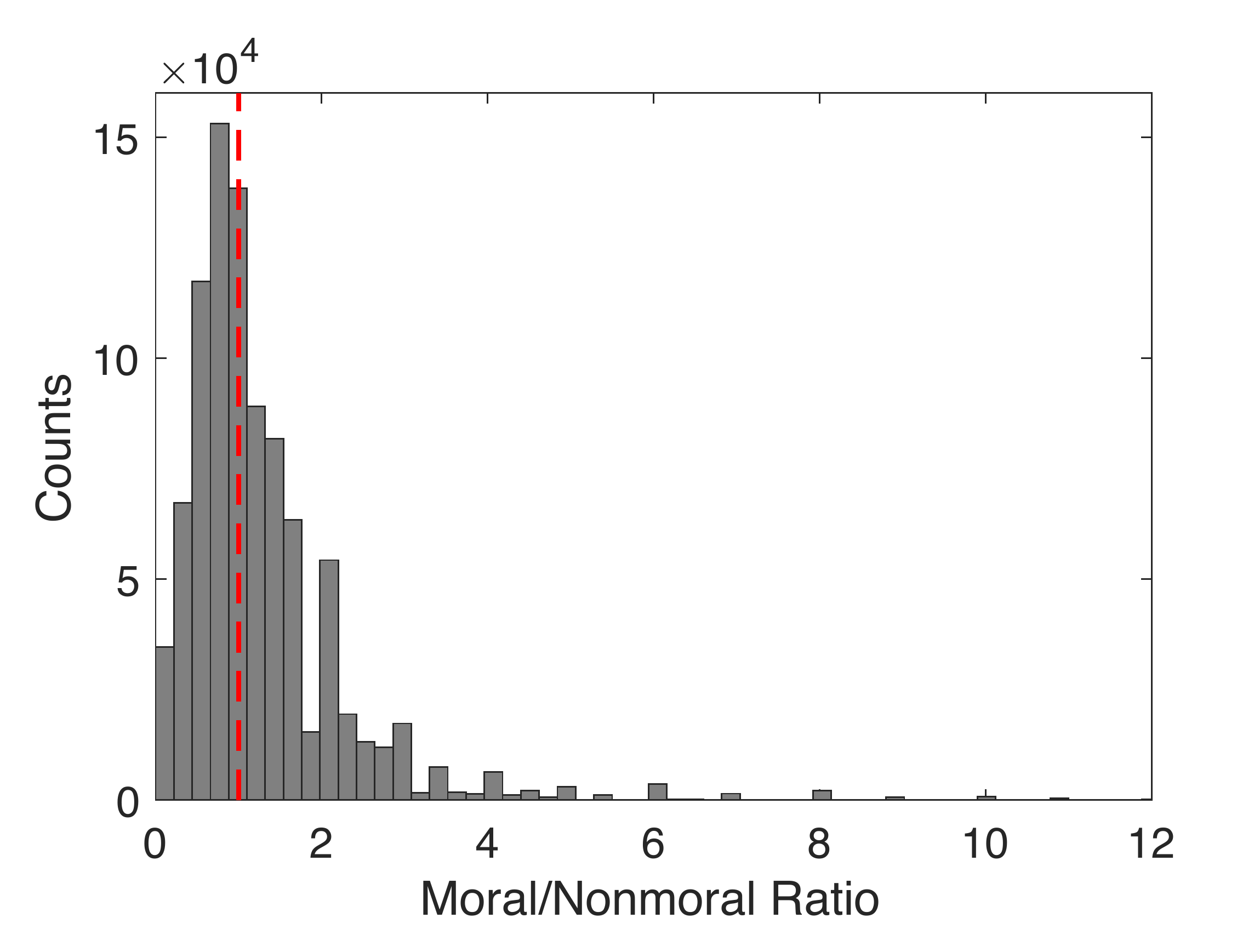}
    \caption{Histogram of the ratio of moral words to non-moral words in Syrian White Helmets-related tweets to question the necessity of moral foundations analysis and the capability of the EMFD used in this study.}
    \label{fig:mnratio}
\end{figure}

This study aims to quantify the latent moral loadings of White Helmets-related posts of Twitter users and to understand the dynamics of polarity in the moral foundations of users. For this purpose, we used EMFD to calculate the moral scores in five dimensions of MFT. The algorithm gives ten results for each tweet input; these are the values of the five dimensions of moral foundations and five sentiment values to make an inference about being vice or virtue rhetoric of each moral foundation. Suppose that $T^i_t$ is the $i^{th}$ tweet in data set published at time $t$, it may either have a virtue rhetoric in the first moral dimension of Authority/Respect $T^i_t(M_{R1})$ or a vice rhetoric $T^i_t(M_A)$. Same is valid for other dimensions; i.e. $T^i_t(M_C)$ or $T^i_t(M_R)$ for Care/Harm, $T^i_t(M_R2)$ or $T^i_t(M_F)$ for Reciprocity/Fairness, $T^i_t(M_L)$ or $T^i_t(M_I)$ for Loyalty/In-group and $T^i_t(M_S)$ and $T^i_t(M_P)$ for Sanctity/Purity. To question the necessity of moral foundations analysis and the capability of the EMFD used in this study, we first measured the ratio of moral words to non-moral words and recognized that almost 57\% of the tweets include more moral words than non-moral words unsurprisingly (Figure \ref{fig:mnratio}).

\begin{figure}
  \centering
  \includegraphics[width=0.6\linewidth]{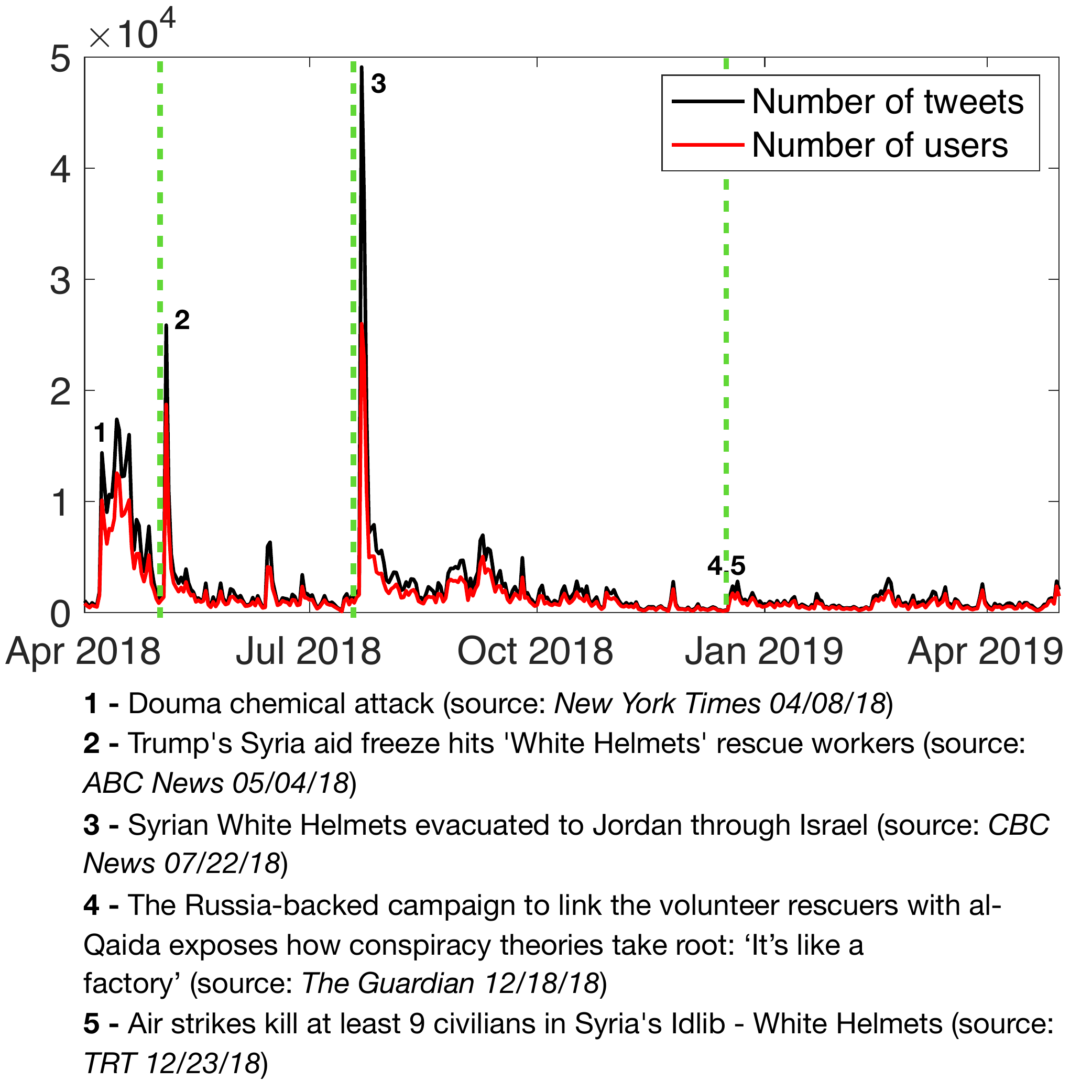}
  \caption{Daily number of user activities (black line) and unique user involved (red line). (Titles of the news possibly related to the bursts in Twitter cascades are given below and shown with green dashed line.}
  \label{fig:daily}
\end{figure}

Later, we examined the daily number of user activities on White Helmets-related discussions and the number of unique users who are involved in these activities from April 1st, 2018 to April 30th, 2019, to obtain a better understanding of the data (Figure \ref{fig:daily}). Unsurprisingly, the long time range of the data set brought instability in daily user activities, and some essential events triggered the burstiness of the specific Twitter cascades at specific times. The first peak in daily Twitter activities is observed just after a chemical attack in Douma occurred on April 7th, 2018. Another important event that brought awareness towards White Helmets and caused another large cascade in Twitter is the suspension of the financial aid of the U.S. to the Syrian humanitarian group in a short time later on May 04th, 2018. The highest peak in the time series of user activities on Twitter, on the other hand, is observed when Israel evacuated White Helmets and their families to Jordan on July 22nd, 2018. The rest of the data also covers multiple events, and relatively lower peaks are observed in the daily activities of Twitter users in response to them.  

The lengthiness of the time period and the diversity of the events in the data set necessitate a more detailed analysis of latent moral foundations rather than applying this method to the entire data set. People's stance might change over time or can be affected by others, or the changes in the ongoing set of events might affect people's attitudes even if there is no external effect of any other source. Therefore, we examined how moral foundations change across time on different White Helmet-related narratives. 

\subsection{Moral Foundations across Time}

\textbf{RQ1:} Does the moral rhetoric of user-generated content on Twitter change over time?

\begin{figure*}
  \centering
  \includegraphics[width=\linewidth]{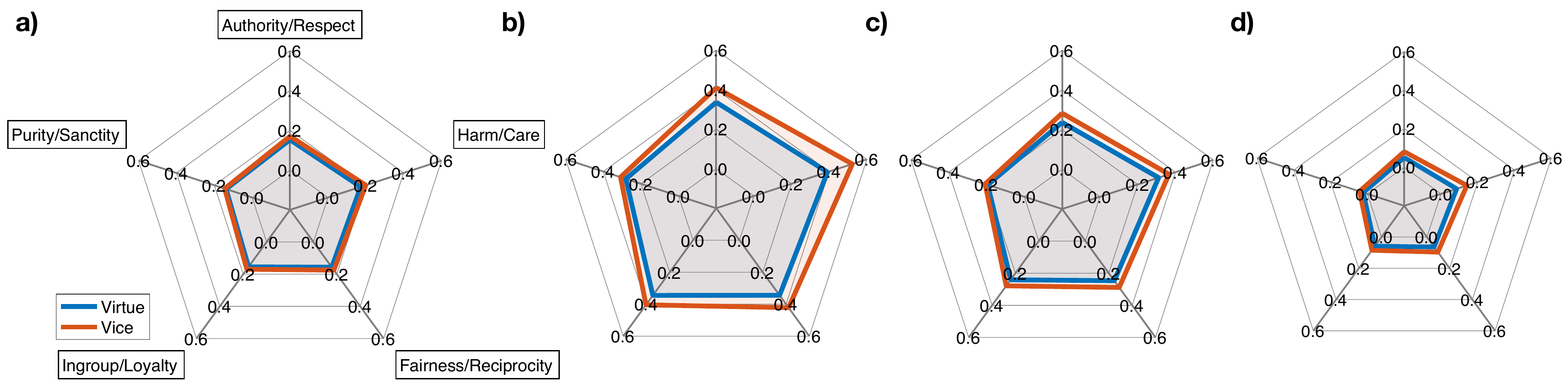}
  \caption{The normalized vice and virtue moral foundations from a) April 1st, 2018 to May 04th, 2018, b) May 04th, 2018 to Jul 22th, 2018, c) Jul 22th, 2018 to Dec 18th, 2018 and d) Dec 18th, 2018 to Apr 30th, 2019.}
   \label{fig:bytime}
\end{figure*}

To understand the change in the moral foundations across time, we explored the moral rhetoric of tweets in terms of vice and virtue moral words represented in EMFD in different time periods. We determined each time range by taking the major events that caused bursts in Twitter activities into account. The first time range $t_1$ covers tweets between the time of Douma chemical attack and Trump's Syria aid freeze (April 1st, 2018 - May 04th, 2018). The second time range $t_2$ starts with the end of the first time period and covers the events until the evacuation of White Helmets to Jordan through Israel (May 04th, 2018 - Jul 22th, 2018). Despite the relative un-burstiness of the rest of the data, third $t_3$ and fourth time ranges $t_4$ are divided before and after becoming the target of a disinformation campaign that positions White Helmets as an al-Qaida-linked terrorist organization (Dec 18th, 2018).

Figure \ref{fig:bytime} shows how the use of vice (Authority, Harm, Fairness, In-group, Purity) and virtue (Respect, Care, Fairness, Loyalty, Sanctity) moral rhetoric has varied in White Helmets-related user activities on Twitter across time. For example, Figure \ref{fig:bytime}.a is obtained as follows:

\begin{equation}
    Authority = \frac{\sum_{t<t_1}{T^i_t(M_A)}}{N(T_{t<t_1 \cap T^i_t(M_A) \neq 0 })}
\end{equation}

Here, $N(T_{t<t_1 \cap T^i_t(M_A) \neq 0})$ is the number of tweets that are classified as having an Authority moral loading score rather than having a Respect moral loading,  and published before $t_1$. Likewise, the value of Respect dimension is obtained as:
\begin{equation}
    Respect = \frac{\sum_{t<t_1}{T^i_t(M_R)}}{N(T_{t<t_1 \cap T^i_t(M_R) \neq 0 })}
\end{equation}
where $N(T_{t<t_1 \cap T^i_t(M_R) \neq 0})$ is the number of tweets that are classified as having an Respect moral loading score rather than having a Authority  moral loading, and published before $t_1$. Since time ranges and number of tweets in each time range greatly differ, values are normalized with the count of tweets in the denominator for a fair comparison. 

Moral loadings in the five dimensions of the vice moral foundations are shown with red lines and blue lines denote the dimensions of the virtue moral loadings. These sub-figures explain how language around certain words and concepts evolves over time in terms of the strength in their rhetoric. Although these figures give no clue about the number of tweets include moral rhetoric in each dimension, we observe that the strength of moral rhetoric of the tweets in the second time period after the evacuation of the White Helmets to Jordan significantly increases ($p<0.005$). We show that moral rhetoric of the tweets in the last time range is the least significant ($p<0.001$), and followed by the tweets in the first, third and second time ranges, in order. In all cases, vice moral rhetoric is found slightly more than the virtue ones in tweets, and this difference is relatively more transparent in the second time range. The most important point is that the pattern of the moral rhetoric of the tweets among five dimensions are very similar while its amount varies by time, i.e. Tweets including Harm/Care moral rhetoric is generally higher in moral loading score, followed by those classified as having an Authority/Respect or a Fairness/Reciprocity moral rhetoric. The least strong moral foundation in terms of moral loading scores in the White Helmet-related tweets is Purity/Sanctity.

\subsection{Moral Foundations across White Helmet Narratives}
\textbf{RQ2:} Does the moral rhetoric of user-generated content on Twitter change across different narratives? 

\begin{figure}
  \centering
  \includegraphics[width=0.5\linewidth]{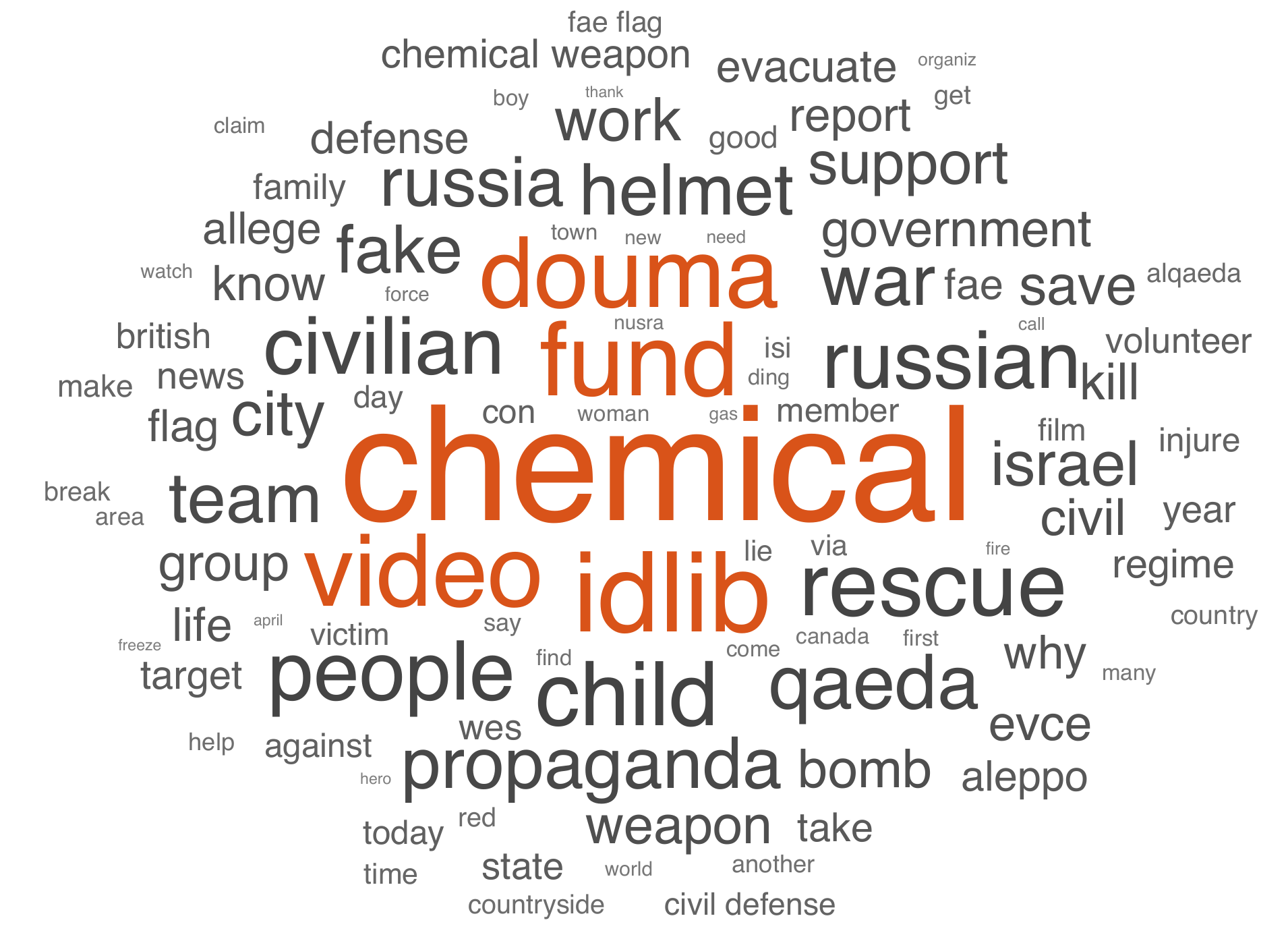}
  \caption{The word cloud of the cleaned White Helmets data set.}
  \label{fig:allword}
\end{figure}

\begin{figure*}
  \centering
  \includegraphics[width=\linewidth]{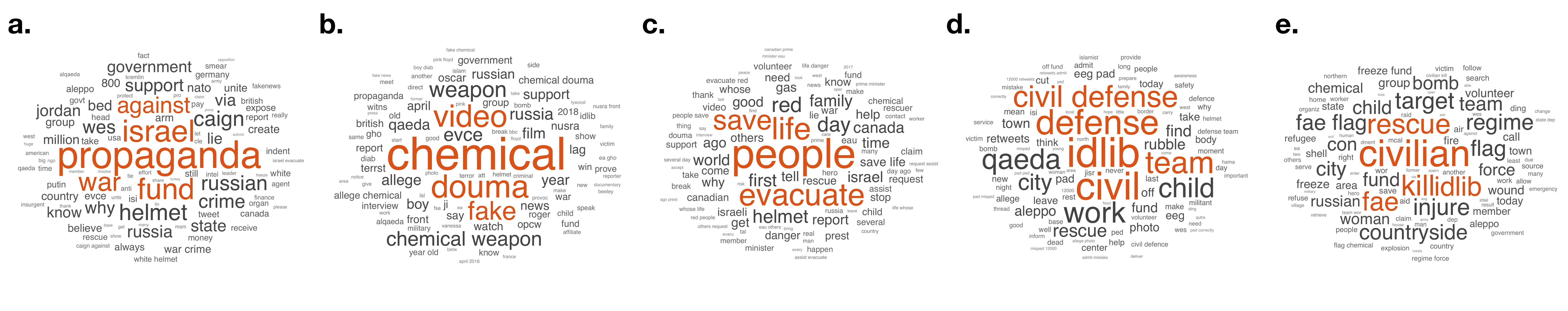}
  \caption{Five major LDA topics of White Helmets-related tweets from April 1st, 2018 to April 30th, 2019.}
  \label{fig:lda}
\end{figure*}

As mentioned before, we discarded some most frequent words in the data set to prevent a masking effect on others (e.g. White Helmets). The most common one-gram lexicons in data corpus after pre-processing can be seen in Figure \ref{fig:allword}. Chemical, video, Idlib, Douma and fund are the most frequently used five words in the White Helmets-related tweets. Since these words belong to the tweets in response to different events, we investigated how the use of moral rhetoric has shifted in different event-related topics in Twitter. Therefore, we used the LDA algorithm to find the topic clusters in the White Helmets data set. The LDA analysis using stochastic approximate variational Bayes solver gave us the optimum coherence value when the number of topics $(K)$ is equal to 5, i.e., the coherence values significantly increased with increasing number of topics until $K=5$. After that, the increase in the coherence value becomes very marginally or the coherence value fluctuates with increasing number of topics. Since we want to focus on major topics rather than all variants, we kept number of topics relatively low. Figure \ref{fig:lda} shows the five major word clusters (one-gram and bi-gram) in the results obtained from LDA. As expected, words related to the major events caused bursts in user activities in Figure \ref{fig:daily} are scattered on different clusters and topics are shaped around them. While the first cluster $(tpc_1)$ mainly captures the tweets related to the event of Syrian White Helmets evacuation to Jordan through Israel, it also covers tweets linked with Russian propaganda against them. The second topic cluster $(tpc_2)$, on the other hand, includes the words in the tweets related more to the fake videos during and after the Douma chemical attack. Third cluster $(tpc_3)$ contains tweets with general stance towards White Helmets, them being volunteers, saving people's lives etc. Fourth $(tpc_4)$ and fifth topic clusters $(tpc_5)$ are somewhat tangled and mainly covers tweets about the Idlib air strike.

\begin{figure}
  \centering
  \includegraphics[width=0.5\linewidth]{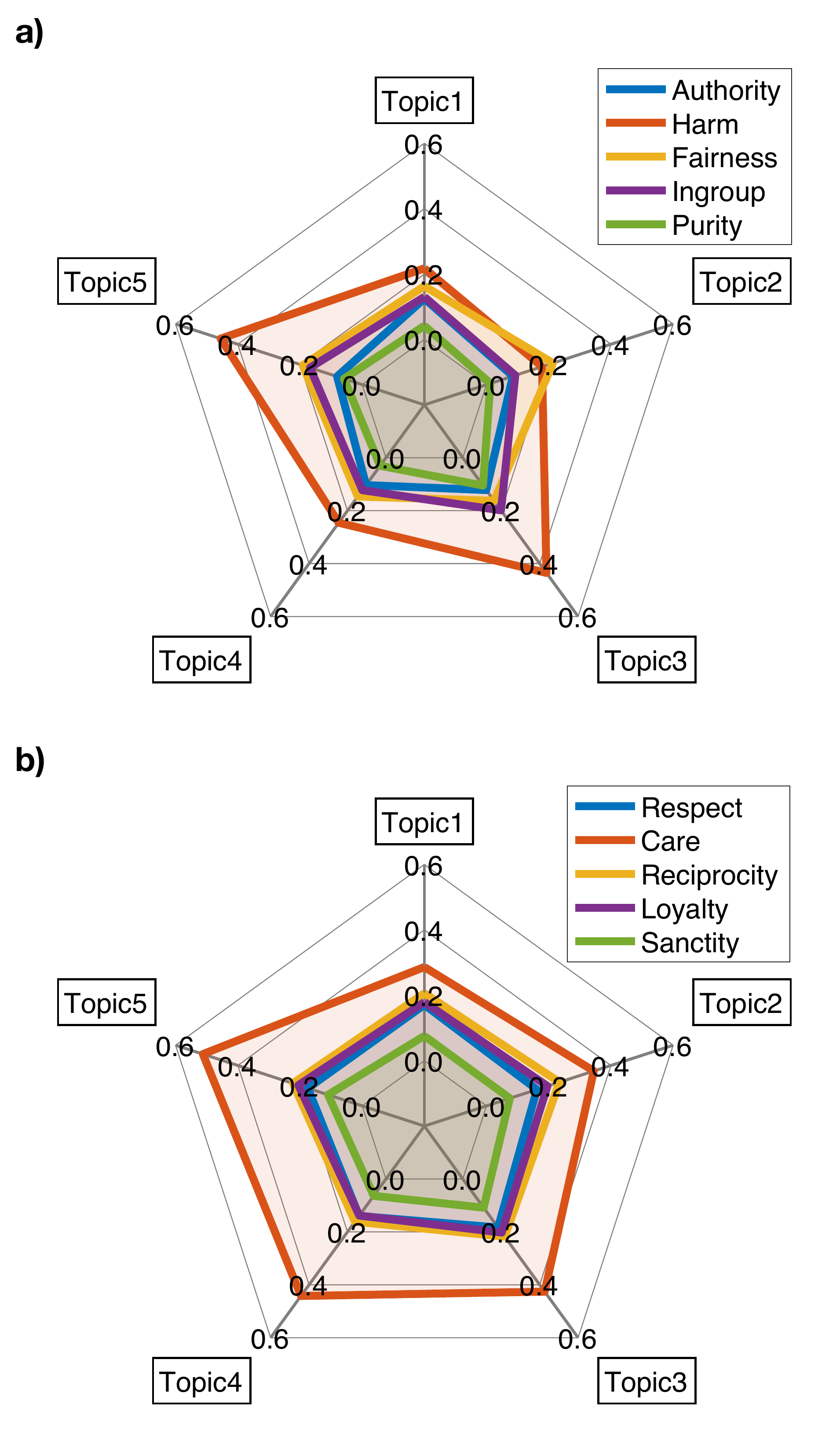}
  \caption{Five a) vice, b) virtue moral foundations across five topics in White Helmets-related tweets from April 1st, 2018 to April 30th, 2019.}
  \label{fig:moraltopic}
\end{figure}

Since people's stances and the corresponding language use in their tweets might show a different moral rhetoric with respect to the narrative behind the event, we investigated the diversity in the MFT dimensions across five main narratives found in data. Figure \ref{fig:moraltopic} shows how negative/vice (Figure  \ref{fig:moraltopic}.a) and positive/virtue (Figure  \ref{fig:moraltopic}.b) emotions/stances vary across five major topics in Figure \ref{fig:lda}. To obtain these results, we first assigned topics to each tweet by computing the maximum value in topic mixtures that is given as one of the output of LDA, and calculated the moral loadings for each five dimensions for virtue and vice language separately. Authority/Respect dimensions related to $tpc_1$, for example, obtained as follows:

\begin{equation}
    Authority = \frac{\sum_{T^i \in tpc_1}{T^i(M_A)}}{N(T^i \in tpc_1 \cap T^i(M_A) \neq 0)}
\end{equation}

Here, $N(T^i \in tpc_1 \cap T^i(M_A) \neq 0)$ is the number of tweets that has score in Authority dimension and mostly related to $tpc_1$. Likewise, the value of Respect dimension is obtained as:

\begin{equation}
    Respect = \frac{\sum_{T^i \in tpc_1}{T^i(M_R)}}{N(T^i \in tpc_1 \cap T^i(M_A) \neq 0)}
\end{equation}
where $N(T^i \in tpc_1 \cap T^i(M_R) \neq 0)$ is the number of tweets that has score in Respect dimension and mostly related to $tpc_1$. Since number of tweets in each topic clusters greatly differ, values are normalized with the count of tweets in the denominator for a fair comparison. 

\begin{figure*}
  \centering
  \includegraphics[width=\linewidth]{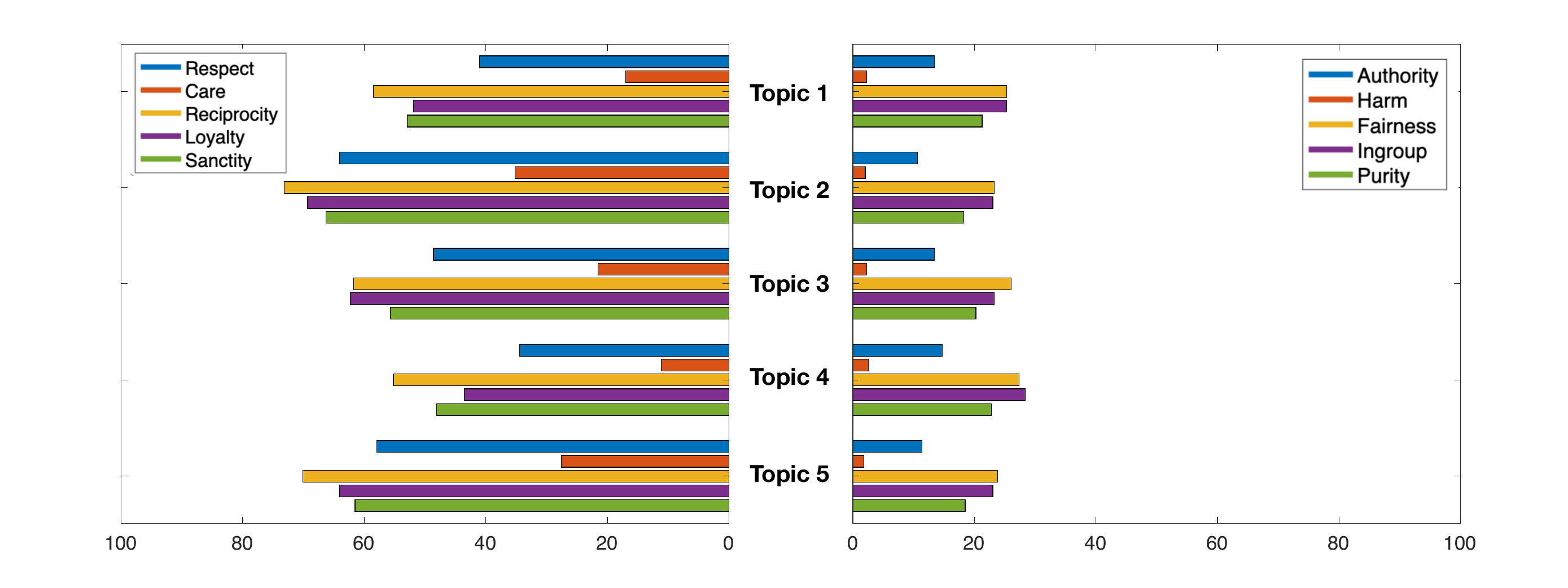}
  \caption{The percentage of days in which tweets related to the relative topics have vice and virtue moral loading.}
  \label{fig:toppol}
\end{figure*}

Figure \ref{fig:moraltopic}.a shows the strength of the five vice moral dimensions used in the tweets across five topics, while Figure \ref{fig:moraltopic}.a represents five virtue moral dimensions across them. The abundance of Harm/Care rhetoric across all topics is noteworthy and it is more significant in virtue tweets. Major exceptions are i) the strength of the morality value of 'Fairness' tweets of Topic 2 that exceeds the morality value of 'Harm' tweets, ii) Average moral loading values are very close to each other in Topic 1. Despite of the small differences, people tend to use more Harm/Care rhetoric associated with White Helmets-related events, followed by Fairness/Reciprocity, In-group/Loyalty and Authority/Respect rhetoric. Among both vice and virtue moral loadings, tweets with Purity/Sanctity rhetoric are observed as the weakest morality dimension in terms of their morality loading scores. 

\subsection{Polarization of Morality}

\textbf{RQ3:} In which dimension of the MFT people are more polarized in terms of their moral loadings?\\
\textbf{RQ4:} Do morality dimensions show similar pattern, or independent from each other? 

\begin{figure*}
  \centering
  \includegraphics[width=\linewidth]{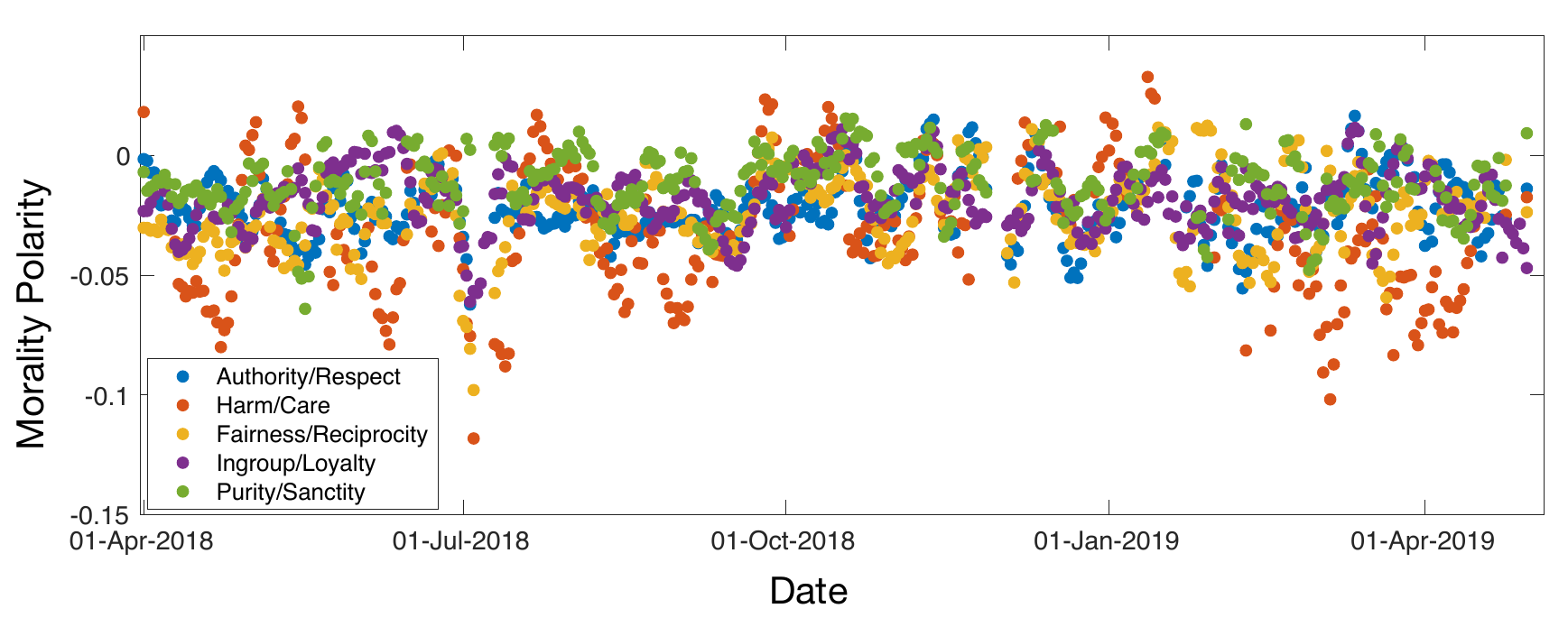}
  \caption{The daily average of morality polarization in each dimension of MFT.}
  \label{fig:morpol}
\end{figure*}

It is important to understand how differences in morality vary and result in morality polarization, as well as understanding the polarity in each dimension of MFT evolve independently or display similar patterns. In all of the previous analyses, we investigated the strength of moral loadings in each dimension of MFT by considering the moral loadings of each tweet. To make a better inference about the morality polarization, we first computed the percentage of the number of days in which vice and virtue tweets exist in moral dimensions of five different topics. Since there are only a few days in the data set that lack a moral rhetoric in tweets, the summation of the vice and virtue rhetoric in individual bar-pairs adds up approximately to \%100. When we observe the quantity of the tweets with a moral rhetoric, we see that people tend to share more tweets involving the virtue moral rhetoric than the vice rhetoric in their tweets. Even though there is no significant difference between the use of Fairness/Reciprocity, In-group/Loyalty or Purity/Sanctity rhetoric, the less use of Harm/Care rhetoric is significant and remarkable (Figure \ref{fig:toppol}). 

We distinguished the quantity of the tweets with each moral rhetoric; however, the qualities (value of the moral loading) may differ in those tweets. Therefore, we calculated the morality polarization by taking the difference of the averages of vice and virtue moral loadings. The morality polarization in Authority/Respect at time $t^*$ for example, is calculated as:

\begin{equation}
    Authority/Respect = \frac{\sum_{t=t^*}{T^i_t(M_R)}}{N(T_{t=t^* \cap T^i_t(M_R) \neq 0})}-\frac{\sum_{t=t^*}{T^i_t(M_A)}}{N(T_{t=t^* \cap T^i_t(M_A) \neq 0})}
\end{equation}

\begin{table*}
  \centering
  \caption{Shannon Entropy Values of CRQA on Moral Foundations Dimensions}
  \label{tab:ent}
  \begin{tabular}{cccccl}
    \toprule
   &Authority&Harm&Fairness&In-group&Purity\\
    \midrule
    Authority/Respect&&0.2221&0.2683&0.3014&0.2091\\
    Harm/Care&&&0.2392&0.3084&0.2632 \\
    Fairness/Reciprocity&&&&0.1734&0.2006 \\
    In-group/Loyalty&&&&&0.1998 \\
  \bottomrule
\end{tabular}
\end{table*}

Having negative polarity scores once again show the abundance of virtue rhetoric compared to the vice rhetoric in tweets with moral loadings. The color code of the outliers show that the morality polarization is mostly observed in Harm/Care tweets. We can easily realize morality polarity has some common patterns, co-rises and co-falls in some dimensions. To understand the correlations and recurrence of these patterns in time series of each moral foundations, we applied CRQA to each pairwise time series. Shannon entropy, which is one of the most common performance metric measures of the recurrence plots, are given in Table \ref{tab:ent}. High entropy denotes more uncertainty in the co-aggregation of pairwise time series of morality polarization. The minimum Shannon Entropy value is observed in the CRQA between time series of morality polarization in Fairness/Reciprocity and In-group/Loyalty, i.e., the increase and decrease in the morality polarization in tweets with Fairness/Reciprocity and In-group/Loyalty show a  similar pattern. Daily morality polarization values show more distinct patterns between tweets with In-group/Loyalty and Authority/Respect and In-group/Loyalty and Harm/Care. It should be noted that Shannon entropy on the diagonal of CRQA plots gives the correlation between two time series, not  causality. A causality analysis is required to consider the time lags between patterns.

\section{Conclusion}
Figuring out the moral rhetoric in textual data is crucial for understanding the latent dynamics of people's cognitive processes. Specifically for topics that are sensitive in nature, such as politics, environmental and societal issues etc., relying solely on sentiment analyses may prove ineffective due to hidden motives and unobservable intentions that are extant in the data structure. Adding the morality component to the sentiment analysis therefore makes a contribution to the literature on topic analysis and narrative extraction by highlighting latent intentions. In this study, we apply the Moral Foundations Theory to the case of Syrian White Helmets related Twitter misinformation data to have a better understanding of how the influence machine works, and thus pave the way for a better strategy to combat misinformation in general. To quantify the dimensions of morality in Twitter narratives, we use the Extended Moral Foundations Dictionary to investigate the change of moral dimensions and polarization of 
morality. Our results indicate that the pattern of moral dimensions in our data set remains unchanged across the five dimensions for the 13-month time period. Still, significant events may trigger an increase in the strength of tone involving any moral rhetoric as people become more sensitive and opinionated. In terms of tweet sharing patterns, it has been shown that people tend to share more tweets involving the virtue moral rhetoric than the tweets involving the vice rhetoric. Although the number of tweets involving the Harm/Care dimension is the least among all dimensions, these tweets come out strong in terms of including more words related to harm/care. Also, polarization of morality is the most prominent among Harm/Care related tweets. As far as patterns go, the increase and the decrease in the polarization of morality in tweets involving Fairness/Reciprocity and In-group/Loyalty display similar patterns.

\section{Acknowledgment}
This work is partially supported by grant FA8650-18-C-7823 from the Defense Advanced Research Projects Agency (DARPA).

\bibliographystyle{unsrt}  
\bibliography{references}  

\begin{thebibliography}{10}

\bibitem{parry2004ancient}
Richard Parry.
\newblock Ancient ethical theory.
\newblock 2004.

\bibitem{graham2013moral}
Jesse Graham, Jonathan Haidt, Sena Koleva, Matt Motyl, Ravi Iyer, Sean~P
  Wojcik, and Peter~H Ditto.
\newblock Moral foundations theory: The pragmatic validity of moral pluralism.
\newblock In {\em Advances in experimental social psychology}, volume~47, pages
  55--130. Elsevier, 2013.

\bibitem{garibay2019polarization}
Ivan Garibay, Alexander~V Mantzaris, Amirarsalan Rajabi, and Cameron~E Taylor.
\newblock Polarization in social media assists influencers to become more
  influential: analysis and two inoculation strategies.
\newblock {\em Scientific Reports}, 9(1):1--9, 2019.

\bibitem{AdvNar}
Ben Decker.
\newblock {\em Adversarial Narratives: A New Model for Disinformation}, 2019
  (accessed March 9, 2019).

\bibitem{rezapour2019enhancing}
Rezvaneh Rezapour, Saumil~H Shah, and Jana Diesner.
\newblock Enhancing the measurement of social effects by capturing morality.
\newblock In {\em Proceedings of the Tenth Workshop on Computational Approaches
  to Subjectivity, Sentiment and Social Media Analysis}, pages 35--45, 2019.

\bibitem{mutlu2020degree}
Ece~C Mutlu and Ivan Garibay.
\newblock The degree-dependent threshold model: Towards a better understanding
  of opinion dynamics on online social networks.
\newblock {\em arXiv preprint arXiv:2003.11671}, 2020.

\bibitem{reisenzein2009emotions}
Rainer Reisenzein.
\newblock Emotions as metarepresentational states of mind: Naturalizing the
  belief--desire theory of emotion.
\newblock {\em Cognitive Systems Research}, 10(1):6--20, 2009.

\bibitem{whitley2008cognition}
David~S Whitley.
\newblock Cognition, emotion, and belief: First steps in an archaeology of
  religion.
\newblock {\em Belief in the past: Theoretical approaches to the archaeology of
  religion}, pages 85--103, 2008.

\bibitem{paulus2012emotion}
Martin~P Paulus and J~Yu Angela.
\newblock Emotion and decision-making: affect-driven belief systems in anxiety
  and depression.
\newblock {\em Trends in cognitive sciences}, 16(9):476--483, 2012.

\bibitem{pennebaker2001linguistic}
James~W Pennebaker, Martha~E Francis, and Roger~J Booth.
\newblock Linguistic inquiry and word count: Liwc 2001.
\newblock {\em Mahway: Lawrence Erlbaum Associates}, 71(2001):2001, 2001.

\bibitem{graham2009liberals}
Jesse Graham, Jonathan Haidt, and Brian~A Nosek.
\newblock Liberals and conservatives rely on different sets of moral
  foundations.
\newblock {\em Journal of personality and social psychology}, 96(5):1029, 2009.

\bibitem{hofmann2014morality}
Wilhelm Hofmann, Daniel~C Wisneski, Mark~J Brandt, and Linda~J Skitka.
\newblock Morality in everyday life.
\newblock {\em Science}, 345(6202):1340--1343, 2014.

\bibitem{feinberg2013moral}
Matthew Feinberg and Robb Willer.
\newblock The moral roots of environmental attitudes.
\newblock {\em Psychological science}, 24(1):56--62, 2013.

\bibitem{clifford2013words}
Scott Clifford and Jennifer Jerit.
\newblock How words do the work of politics: Moral foundations theory and the
  debate over stem cell research.
\newblock {\em The Journal of Politics}, 75(3):659--671, 2013.

\bibitem{kaur2016quantifying}
Rishemjit Kaur and Kazutoshi Sasahara.
\newblock Quantifying moral foundations from various topics on twitter
  conversations.
\newblock In {\em 2016 IEEE International Conference on Big Data (Big Data)},
  pages 2505--2512. IEEE, 2016.

\bibitem{hoover2019moral}
Joe Hoover, Gwenyth Portillo-Wightman, Leigh Yeh, Shreya Havaldar,
  Aida~Mostafazadeh Davani, Ying Lin, Brendan Kennedy, Mohammad Atari, Zahra
  Kamel, Madelyn Mendlen, et~al.
\newblock Moral foundations twitter corpus: A collection of 35k tweets
  annotated for moral sentiment.
\newblock {\em Social Psychological and Personality Science}, page
  1948550619876629, 2019.

\bibitem{sagi2014moral}
Eyal Sagi and Morteza Dehghani.
\newblock Moral rhetoric in twitter: A case study of the us federal shutdown of
  2013.
\newblock In {\em Proceedings of the Annual Meeting of the Cognitive Science
  Society}, volume~36, 2014.

\bibitem{starbird2018ecosystem}
Kate Starbird, Ahmer Arif, Tom Wilson, Katherine Van~Koevering, Katya Yefimova,
  and Daniel Scarnecchia.
\newblock Ecosystem or echo-system? exploring content sharing across
  alternative media domains.
\newblock In {\em Twelfth International AAAI Conference on Web and Social
  Media}, 2018.

\bibitem{haidt2004intuitive}
Jonathan Haidt and Craig Joseph.
\newblock Intuitive ethics: How innately prepared intuitions generate
  culturally variable virtues.
\newblock {\em Daedalus}, 133(4):55--66, 2004.

\bibitem{sagi2014measuring}
Eyal Sagi and Morteza Dehghani.
\newblock Measuring moral rhetoric in text.
\newblock {\em Social science computer review}, 32(2):132--144, 2014.

\bibitem{araque2020moralstrength}
Oscar Araque, Lorenzo Gatti, and Kyriaki Kalimeri.
\newblock Moralstrength: Exploiting a moral lexicon and embedding similarity
  for moral foundations prediction.
\newblock {\em Knowledge-based systems}, 191:105184, 2020.

\bibitem{garten2016morality}
Justin Garten, Reihane Boghrati, Joe Hoover, Kate~M Johnson, and Morteza
  Dehghani.
\newblock Morality between the lines: Detecting moral sentiment in text.
\newblock In {\em Proceedings of IJCAI 2016 workshop on Computational Modeling
  of Attitudes}, 2016.

\bibitem{hoffman2013stochastic}
Matthew~D Hoffman, David~M Blei, Chong Wang, and John Paisley.
\newblock Stochastic variational inference.
\newblock {\em The Journal of Machine Learning Research}, 14(1):1303--1347,
  2013.

\bibitem{foulds2013stochastic}
James Foulds, Levi Boyles, Christopher DuBois, Padhraic Smyth, and Max Welling.
\newblock Stochastic collapsed variational bayesian inference for latent
  dirichlet allocation.
\newblock In {\em Proceedings of the 19th ACM SIGKDD international conference
  on Knowledge discovery and data mining}, pages 446--454, 2013.

\bibitem{griffiths2004finding}
Thomas~L Griffiths and Mark Steyvers.
\newblock Finding scientific topics.
\newblock {\em Proceedings of the National academy of Sciences}, 101(suppl
  1):5228--5235, 2004.

\bibitem{asuncion2012smoothing}
Arthur Asuncion, Max Welling, Padhraic Smyth, and Yee~Whye Teh.
\newblock On smoothing and inference for topic models.
\newblock {\em arXiv preprint arXiv:1205.2662}, 2012.

\bibitem{teh2007collapsed}
Yee~W Teh, David Newman, and Max Welling.
\newblock A collapsed variational bayesian inference algorithm for latent
  dirichlet allocation.
\newblock In {\em Advances in neural information processing systems}, pages
  1353--1360, 2007.

\bibitem{coco2014cross}
Moreno~I Coco and Rick Dale.
\newblock Cross-recurrence quantification analysis of categorical and
  continuous time series: an r package.
\newblock {\em Frontiers in psychology}, 5:510, 2014.

\bibitem{wallot2019multidimensional}
Sebastian Wallot.
\newblock Multidimensional cross-recurrence quantification analysis (mdcrqa)--a
  method for quantifying correlation between multivariate time-series.
\newblock {\em Multivariate behavioral research}, 54(2):173--191, 2019.

\end{thebibliography}
\end{document}